\documentclass[preprint,showpacs,preprintnumbers,amsmath,amssymb]{revtex4}
\usepackage{graphicx}
\usepackage{dcolumn}
\usepackage{bm}
\usepackage{epsfig}
\usepackage{epstopdf}
\begin{document}
\title{Reformulation of the Georgi-Glashow model and some constraints on its classical fields}
\author{Ahmad Mohamadnejad}\altaffiliation {a.mohamadnejad@ut.ac.ir}
\affiliation{Young Researchers and Elite Club, Islamshahr Branch, Islamic Azad University, Islamshahr, Iran}
\begin{abstract}
We study the SU(2) Georgi-Glashow model and suggest a decomposition for its fields and obtain a Lagrangian based on new variables.
We use Cho's restricted decomposition as a
result of a vacuum condition of the Georgi-Glashow model. This model with no external sources leads us to the Cho extended decomposition.
We interpret the puzzling field, $ \textbf{n} $, in Cho's decomposition as the color direction of the scalar field in the Georgi-Glashow model.
We also study another constraint, condensate phase, and generalize Cho's extended decomposition.
Finally, we argue about a decomposition form that Faddeev and Niemi proposed in this constrained Georgi-Glashow model.

\end{abstract}


\maketitle

\section{Introduction} \label{sec1}

the Georgi-Glashow model is a  grand unification theory (GUT) proposed in 1974 \cite{Georgi}. In this model the standard model gauge groups
 $ SU(3) \times SU(2) \times U(1) $ are combined into a single simple gauge group-SU(5). The unified group SU(5) is then thought to be spontaneously broken to the standard model subgroup at some high-energy scale called the grand unification scale. On the other hand, the modern era of the monopole theory started in 1974, when 't Hooft and Polyakov independently discovered monopole solutions of the SU(2) Georgi-Glashow model \cite{'t Hooft1,Polyakov1}.
The essence of this breakthrough is that while a Dirac monopole could be incorporated in an Abelian theory, some non-Abelian models, like
the Georgi-Glashow model, inevitably contain monopolelike solutions.

Monopoles can explain quark confinement via the dual Meissner effect in 
Yang-Mills theories like quantum chromodynamics \cite{Nambu,'t Hooft2,Mandelstam}.
Unlike the Georgi-Glashow model, there is no scalar matter field in
 Yang-Mills theories. Moreover, it is believed that the ultraviolet and infrared limits of a Yang-Mills theory
represent different phases \cite{'t Hooft3,Polyakov2}. Perturbative methods are appropriate for the
ultraviolet limit where the Yang-Mills theory is asymptotically free, but
for the infrared limits the Yang-Mills theory becomes strongly coupled and
the perturbative technique fails. In this regime nonperturbative methods
must be developed.

One of the nonperturbative approaches is the decomposition of the Yang-Mills
field to some other variables more appropriate for the low-energy limit \cite{Cho1,Cho2,FN1,FN2,FN3,FN4,Shabanov1,Shabanov2}.
This method proposed by Cho \cite{Cho1} and developed by Faddeev
and Niemi \cite{FN1,FN2,FN3,FN4}, and Shanbanov \cite{Shabanov1,Shabanov2}. Cho introduced an additional
magnetic symmetry which leads to a decomposition for the Yang-Mills field
with four dynamical degrees of freedom \cite{Cho1}. Based on this decomposition,
he tried to construct a local Lagrangian field theory of the monopole 
exhibiting a duality between the electric and the magnetic charges.
He also extended his decomposition in order to contain all degrees of freedom of the SU(2)
Yang-Mills field \cite{Cho2}. Faddeev and Niemi generalized the Cho restricted decomposition to arrive at a dual picture of the
Yang-Mills theory, with the high-energy limit described by a massless and
pointlike transverse polarization of $ A_{\mu} $ and the low-energy limit described
by massive solitonic 
flux tubes which close on themselves in a stable knotlike
configuration.
For a review of the reformulations of Yang-Mills theory, see \cite{Kondo}.

Unlike the Cho restricted decomposition which describes a Yang-Mills field with four
degrees of freedom, the Faddeev-Niemi decomposition enjoys all six dynamical
degrees of freedom of the SU(2) Yang-Mills. Faddeev and Niemi also obtain
the Euler-Lagrange equations of their new action by performing the variation
to new variables. These equations were all proportional to the Yang-Mills
equation and they asserted that their decomposition is complete \cite{FN1}.
However, the Faddeev-Niemi reformulation is inequivalent to Yang-Mills,
but instead describes Yang-Mills coupled to a particular choice of external charge
 \cite{Evslin,Niemi}. Furthermore, there are solutions of the Yang-Mills equation with a
covariantly constant source term that are not solutions to the Faddeev-Niemi
equations \cite{Niemi}.

In this paper, after reviewing Cho-Faddeev-Niemi decomposition of the SU(2) Yang-Mills field, we study 
the Georgi-Glashow model which is a Yang-Mills-Higgs theory with the Higgs field as
the source of external charge.
In this model, there are two fields: the scalar field and the Yang-Mills field. We reformulate this theory based
on four new fields, and we show that this reformulated theory is equivalent to the Georgi-Glashow model at least at the classical level.
Performing variations with respect to these four new variables, we get four Euler-Lagrange equations, one which is trivial and one that is
derivable from the other two. Therefore, there are only two independent equations of motion that are equivalent to the original Georgi-Glasow equations.
One can show that both the Cho restricted and the extended Lagrangians are the limits of this reformulated Georgi-Glashow model by considering some constraints on the classical fields.
A solution
for the Georgi-Glashow model, known as the 't Hooft-Polyakov monopole solution, was presented by 't Hooft and Polyakov \cite{'t Hooft1,Polyakov1}. To have finite energy, they assumed some boundary conditions
for their solutions. These conditions, called "vacuum-conditions", define the Higgs
vacuum of the system. We generalize these conditions for all spacetime.
One of these vacuum conditions leads to the Cho restricted theory, and the other one leads to the condensate phase of the theory.
In the Cho restricted case, the external charge decouples from the model, and in the condensate phase, there is also no external charge, but the new vector field
would be massive.
We also show that the constraint of no external charge leads to either the Cho restricted or the extended decomposition.
In addition, we study a special form of the Faddeev-Niemi reparametrization in the condensate phase and generalize their Lagrangian so that
it obviously contains external charges.

This paper is organized as follows: in Sec. \ref{sec2} we review the Cho-Faddeev-Niemi decomposition.
In Sec. \ref{sec3} we reformulate the Georgi-Glashow model based on four new variables, and we derive the Eueler-Lagrange equations.
In Sec. \ref{sec4} we study some constraints on the classical fields of the reformulated model. In this section we consider the vacuum-conditions
and show that one of them makes the external charge decoupled from the
theory and leads to the Cho restricted Lagrangian.
No external charge condition leads to the Cho extended decomposition.
By considering the other vacuum condition, the condensate phase, we construct an effective Lagrangian which is a generalization of the
Cho-Faddeev-Niemi Lagrangian coupled to an external charge explicitly. Finally,
our conclusion comes in Sec. \ref{sec5}.

\section{Cho-Faddeev-Niemi decomposition of the SU(2) Yang-Mills field} \label{sec2}

The Cho decomposition was introduced a long time ago in an attempt to demonstrate
the monopole condensation in QCD \cite{Cho1,Cho2}. One can obtain the quark confinement potential
using the restricted part of the Cho decomposition via the dual superconductor mechanism in which monopole condensation plays an essential role \cite{M}.

In the Cho decomposition of the Yang-Mills field, an isotriplet unit vector field $ \textbf{n} $,
which selects the Abelian direction at each spacetime point, is introduced. The
Yang-Mills field is restricted to the potential $ \widehat{\textbf{A}}_{\mu} $ which leaves $ \textbf{n} $ invariant,
\begin{equation}
\widehat{\textbf{A}}_{\mu} = A_{\mu} \textbf{n} + \frac{1}{g} \partial_{\mu} \textbf{n} \times \textbf{n}, \label{eq1}
\end{equation}
where $ A_{\mu} = \widehat{\textbf{A}}_{\mu} . \textbf{n} $ and $ \textbf{n} . \textbf{n} = 1 $.
The above decomposition was originally obtained by the following condition,
\begin{equation}
\widehat{\triangledown}_{\mu} \textbf{n} = \partial_{\mu} \textbf{n} + g \widehat{\textbf{A}}_{\mu} \times \textbf{n} = 0,  \label{eq2}
\end{equation}
which means that the restricted Yang-Mills field is the field which leaves $ \textbf{n} $ 
invariant under the parallel transport. In low-energy limit, $ \widehat{\textbf{A}}_{\mu} $ dominates and
has a dual structure. In fact, the field strength tensor $  \widehat{\textbf{F}}_{\mu\nu}  $ made of the
restricted potential $ \widehat{\textbf{A}}_{\mu} $ is decomposed into the electric field strength tensor
$ F_{\mu\nu} $ and magnetic field strength tensor $ H_{\mu\nu} $:
\begin{equation}
\widehat{\textbf{F}}_{\mu\nu} = \partial_{\mu} \widehat{\textbf{A}}_{\nu} -  \partial_{\nu} \widehat{\textbf{A}}_{\mu} + 
g \widehat{\textbf{A}}_{\mu} \times \widehat{\textbf{A}}_{\nu} = (F_{\mu\nu} + H_{\mu\nu})  \textbf{n} ,  \label{eq3}
\end{equation}
where
\begin{eqnarray}
F_{\mu\nu} &=& \partial_{\mu} A_{\nu} - \partial_{\nu} A_{\mu} \nonumber ,\\
H_{\mu\nu} &=& - \frac{1}{g} \textbf{n} . (\partial_{\mu} \textbf{n} \times \partial_{\nu} \textbf{n} ). \label{eq4}
\end{eqnarray}

Note that singularities of $ \textbf{n} $ define $ \pi_{2} (S^{2}) $ which describes the non-Abelian
monopoles. Indeed, one can obtain the Wu-Yang monopole by choosing $ A_{\mu} = 0 $
and $ \textbf{n} = \frac{\textbf{r}}{r} $ \cite{Wu,Cho1}.
Besides, with the $ S^{3} $ compactification of
$ R^{3} $, $ \textbf{n} $ characterizes the Hopf invariant $ \pi_{3} (S^{2}) \simeq \pi_{3} (S^{3}) $
which describes the topologically distinct vacua \cite{Cho3,Woo}.
This indicates that the restricted gauge theory made of $ \widehat{\textbf{A}}_{\mu} $ could
describe the dual dynamics which should play an essential
role in SU(2) QCD, which displays the full topological characters
of the non-Abelian gauge theory.

The restricted potential $ \widehat{\textbf{A}}_{\mu} $ has four degrees of freedom, two for $ A_{\mu} $,
corresponding to two polarizations, and two for $ \textbf{n} $. Although these four
degrees of freedom play an essential role in the infrared limit, but one can
extend the Yang-Mills field as follows:
\begin{equation}
\textbf{A}_{\mu} = \widehat{\textbf{A}}_{\mu} + \textbf{X}_{\mu} , \label{eq5}
\end{equation}
where $  \textbf{X}_{\mu} . \textbf{n} = 0 $.

Under the infinitesimal gauge transformation,
\begin{eqnarray}
\delta \textbf{n} &=& - \textbf{a} \times \textbf{n} \nonumber ,\\
\delta \textbf{A}_{\mu} &=& \frac{1}{g} \triangledown_{\mu} \textbf{a} , \label{eq6}
\end{eqnarray}
one has
\begin{eqnarray}
\delta A_{\mu} &=& \frac{1}{g} \textbf{n} . \partial _{\mu} \textbf{a}  \nonumber ,\\
\delta \widehat{\textbf{A}}_{\mu} &=& \frac{1}{g} \widehat{\triangledown}_{\mu} \textbf{a} \nonumber ,\\
\delta \textbf{X}_{\mu} &=& - \textbf{a} \times\textbf{X}_{\mu} . \label{eq7}
\end{eqnarray}
This shows that $ \widehat{\textbf{A}}_{\mu} $ by itself describes an SU(2) connection
which enjoys the full SU(2) gauge degrees of freedom. Furthermore,
$ \textbf{X}_{\mu} $ transforms covariantly under the gauge transformation.
This confirms that the Cho extended decomposition provides a
gauge-independent decomposition of the non-Abelian potential
into the restricted part $ \widehat{\textbf{A}}_{\mu} $ and the gauge covariant part $ \textbf{X}_{\mu} $.

Using Eq. (\ref{eq3}) and Eq. (\ref{eq5}), we have
\begin{equation}
\textbf{F}_{\mu\nu} = \widehat{\textbf{F}}_{\mu\nu} +  \widehat{\triangledown}_{\mu} \textbf{X}_{\nu} -  \widehat{\triangledown}_{\nu} \textbf{X}_{\mu}
+ g \textbf{X}_{\mu} \times \textbf{X}_{\nu} ,  \label{eq8}
\end{equation}
where
\begin{equation}
\textbf{n} \, . \, \textbf{X}_{\mu} = 0 \quad \Rightarrow \quad \textbf{n} \, . \, \widehat{\triangledown}_{\mu} \textbf{X}_{\nu} = 0 , \label{eq9}
\end{equation}
and the Yang-Mills Lagrangian is
\begin{eqnarray}
L &=& - \frac{1}{4} \textbf{F}_{\mu\nu} \, . \, \textbf{F}^{{\mu\nu}} \nonumber \\
 &=& - \frac{1}{4} \widehat{\textbf{F}}_{\mu\nu} \, . \, \widehat{\textbf{F}}^{\mu\nu} - \frac{1}{4} ( \widehat{\triangledown}_{\mu} \textbf{X}_{\nu} -  \widehat{\triangledown}_{\nu} \textbf{X}_{\mu}).( \widehat{\triangledown}^{\mu} \textbf{X}^{\nu} -  \widehat{\triangledown}^{\nu} \textbf{X}^{\mu}) \nonumber \\
 & & - \frac{g^{2}}{4} (\textbf{X}_{\mu} \times \textbf{X}_{\nu}).(\textbf{X}^{\mu} \times \textbf{X}^{\nu}) 
 - \frac{g}{2} \widehat{\textbf{F}}_{\mu\nu} \, . \, (\textbf{X}^{\mu} \times \textbf{X}^{\nu})   . \label{eq10}
\end{eqnarray}
This shows that the Yang-Mills theory can be viewed as the restricted gauge
theory made of the dual potential $ \widehat{\textbf{A}}_{\mu} $, which has the valence
gluon $ \textbf{X}_{\mu} $ as its source.

The equations of motion that one obtains from the Cho Lagrangian
by varying $ A_{\mu} $ , $ \textbf{X}_{\mu} $ , and $ \textbf{n} $ are given by
\begin{eqnarray}
\partial_{\mu} (F^{\mu\nu}+H^{\mu\nu}+X^{\mu\nu}) &=& - g \textbf{n} . [\textbf{X}_{\mu} \times ( \widehat{\triangledown}^{\mu} \textbf{X}^{\nu} -  \widehat{\triangledown}^{\nu} \textbf{X}^{\mu}) ] ,  \label{eq11} \\
\widehat{\triangledown}_{\mu} ( \widehat{\triangledown}^{\mu} \textbf{X}^{\nu} -  \widehat{\triangledown}^{\nu} \textbf{X}^{\mu}) &=& g (F^{\mu\nu}+H^{\mu\nu}+X^{\mu\nu})
\textbf{n} \times \textbf{X}_{\mu}  , \label{eq12}
\end{eqnarray}
where
\begin{equation}
X^{\mu\nu} = g \textbf{n} . (\textbf{X}^{\mu} \times \textbf{X}^{\nu})  , \label{eq13}
\end{equation}
Eq. (\ref{eq11}) and Eq. (\ref{eq12})  are obtained by varing $ A_{\mu} $ and $ \textbf{X}_{\mu} $, respectively,
and the variation with respect to
$ \textbf{n} $ does not create any new equation of motion. Therefore, $ \textbf{n} $ 
is not a dynamical variable in the Cho decomposition.

Notice that Eq. (\ref{eq11}) and Eq. (\ref{eq12}) are not independent. Indeed, Eq. (\ref{eq11}) can be obtained from Eq. (\ref{eq12}).
Furthermore, Eq. (\ref{eq11}) and Eq. (\ref{eq12}) are identical to the original Yang-Mills equation:
\begin{equation}
\triangledown_{\mu} \textbf{F}^{\mu\nu} = 0 . \label{eq14}
\end{equation}
So the Cho decomposition does not change the dynamics of QCD at the classical level.

Faddeev and Niemi proposed a special form of $ \textbf{X}_{\mu} $ in the Cho decomposition. In this
proposal, only two of four dynamical degrees of freedom are considered for $ \textbf{X}_{\mu} $. Thus,
it does not describe the full QCD. One can propose the following form for $ \textbf{X}_{\mu} $  which respects the constraint (\ref{eq9}):
\begin{equation}
\textbf{X}_{\mu}  = \frac{\phi_{1}}{g^{2}} \partial_{\mu} \textbf{n} + \frac{\phi_{2}}{g^{2}} \textbf{n}\times\partial_{\mu} \textbf{n} , \label{eq15}
\end{equation}
where  $ \phi_{1} $ and $ \phi_{2} $ are real scalar fields. Therefore, one gets
\begin{equation}
\textbf{A}_{\mu} = A_{\mu} \textbf{n} + \frac{1}{g} \partial_{\mu} \textbf{n} \times \textbf{n} + \frac{\phi_{1}}{g^{2}} \partial_{\mu} \textbf{n} + \frac{\phi_{2}}{g^{2}} \textbf{n}\times\partial_{\mu} \textbf{n} . \label{eq16}
\end{equation}
This is the Faddeev-Niemi decomposition. Note that two field degrees of freedom,
$ \phi_{1} $ and $ \phi_{2} $, are added to the Cho restricted decomposition. Now the variation with respect
to $ \textbf{n} $ creates a new equation of motion \cite{FN1}. Therefore, Faddeev and Niemi interpret
$ \textbf{n} $ as a dynamical field. However, unlike the Cho decomposition, the equations of
motion that one obtains from the Faddeev-Niemi Lagrangian are not equivalent to the
original equations of pure Yang-Mills theory \cite{Evslin,Niemi}. Faddeev and Niemi's main proposal was the
completeness of their decomposition in four dimensions, which has been criticized recently \cite{Evslin,Niemi}.

Using Eq . (\ref{eq16}) in the definition of the SU(2) field strength tensor, $ \textbf{F}_{\mu\nu} = \partial_{\mu} \textbf{A}_{\nu} - \partial_{\nu} \textbf{A}_{\mu} + g \textbf{A}_{\mu} \times \textbf{A}_{\nu} $, one gets the following field strength tensor for the Faddeev-Niemi decomposition,
\begin{eqnarray}
\textbf{F}_{\mu\nu} &=& \lbrace F_{\mu\nu} +  (1-\frac{\phi_{1}^{2} + \phi_{2}^{2} }{g^{2}}) H_{\mu\nu} \rbrace \textbf{n}  \nonumber \\
& & +\frac{1}{g^{2}} (D_{\mu} \phi_{1} \partial_{\nu} \textbf{n} - D_{\nu} \phi_{1} \partial_{\mu} \textbf{n} )  \nonumber \\
& & +\frac{1}{g^{2}} (D_{\mu} \phi_{2} \textbf{n} \times \partial_{\nu} \textbf{n} - D_{\nu} \phi_{2} \textbf{n} \times \partial_{\mu} \textbf{n} ) , \label{17}
\end{eqnarray}
where
\begin{eqnarray}
D_{\mu} \phi_{1} &=& \partial_{\mu} \phi_{1} - g C_{\mu} \phi_{2} , \nonumber\\ 
D_{\mu} \phi_{2} &=& \partial_{\mu} \phi_{2} + g C_{\mu} \phi_{1}. \label{eq18}
\end{eqnarray}
Then the Faddeev-Niemi Lagrangian is
\begin{eqnarray}
L &=& -\frac{1}{4} \textbf{F}_{\mu\nu} \,. \, \textbf{F}^{\mu\nu} \nonumber\\ 
 &=& -\frac{1}{4} F_{\mu\nu} F^{\mu\nu} + \frac{1}{2g^{4}} ( \partial_{\mu} \textbf{n}.\partial_{\nu} \textbf{n} - \eta_{\mu\nu} \partial_{\lambda} \textbf{n}.\partial^{\lambda} \textbf{n} )
(D^{\mu}\varphi)^{\ast}  (D^{\nu}\varphi) \nonumber\\
&& + \frac{i}{2g^{3}} H_{\mu\nu} (D^{\mu}\varphi)^{\ast}  (D^{\nu}\varphi)  - \frac{1}{2} H_{\mu\nu} F^{\mu\nu} (1-\frac{\varphi ^{\ast} \varphi}{g^{2}})  \nonumber\\
&& - \frac{1}{4}  H_{\mu\nu} H^{\mu\nu} (1-\frac{\varphi ^{\ast} \varphi}{g^{2}})^{2} ,  \label{eq19}
\end{eqnarray}
where
\begin{eqnarray} 
\varphi &=& \phi_{1} + i \phi_{2} ,  \nonumber\\ 
D_{\mu}\varphi &=& (\partial_{\mu}  + i g A_{\mu} ) \varphi . \label{eq20}
\end{eqnarray}
The Lagrangian (\ref{eq19}) is invariant under the following local U(1) gauge transformations
\begin{eqnarray}
\varphi & \rightarrow & e^{-i\alpha(x)}  \varphi \nonumber\\ 
A_{\mu} & \rightarrow & A_{\mu} + \frac{1}{g} \partial_{\mu} \alpha(x) \label{eq21}
\end{eqnarray}
It is also invariant under rotations of $ \textbf{n} $ in the three-dimensional internal space that forms the non-Abelian group SO(3), which is a global symmetry.

Performing the variations with respect to new variables $ C_{\mu} $, $ \phi_{1} $, $ \phi_{2} $ and $ \textbf{n} $, one gets the following equations of motion:
\begin{eqnarray}
\textbf{n} . \triangledown_{\nu} \textbf{F}^{\mu\nu} &=& 0 , \nonumber\\ 
 \partial_{\mu} \textbf{n} . \triangledown_{\nu} \textbf{F}^{\mu\nu} &=& 0 , \nonumber\\ 
 (\textbf{n} \times \partial_{\mu} \textbf{n}) . \triangledown_{\nu} \textbf{F}^{\mu\nu} &=& 0 , \nonumber\\ 
 (D_{\mu} \phi_{1} - D_{\mu} \phi_{2} \textbf{n} \times) \triangledown_{\nu} \textbf{F}^{\mu\nu} &=& 0. \label{eq22}
\end{eqnarray}

Faddeev and Niemi obtain the Skyrme-Faddeev Lagrangian from their decomposition \cite{FN1}. One can also obtain the Abelian-Higgs model with
the Nielsen-Olesen vortex solutions from the Faddeev-Niemi Lagrangian \cite{M2}.
This suggests that, at low energies, the physical states of the Yang-Mills theory are topological solitons.
In the next section, we use the decomposition ideas for the Georgi-Glashow model which is a Yang-Mills-Higgs theory with the 
Higgs field as the source of external charge.

\section{Georgi-Glashow model in new variables} \label{sec3}

the SU(2) Georgi-Glashow model which describes the coupled gauge and Higgs field has the following classical Lagrangian,
\begin{equation}
L = \frac{1}{2} \triangledown_{\mu} \bm{\phi} \, . \, \triangledown^{\mu} \bm{\phi} -\frac{1}{4} \textbf{F}_{\mu\nu} \,. \, \textbf{F}^{\mu\nu}  - V(\bm{\phi}) ,
 \label{eq23}
\end{equation}
where
\begin{eqnarray}
\triangledown_{\mu} \bm{\phi}  &=& \partial_{\mu} \bm{\phi}  + g  \textbf{A}_{\mu} \times \bm{\phi}   , \nonumber\\ 
\textbf{F}_{\mu\nu} &=& \partial_{\mu} \textbf{A}_{\nu} - \partial_{\nu} \textbf{A}_{\mu} + g \textbf{A}_{\mu} \times \textbf{A}_{\nu}  , \nonumber\\ 
V(\bm{\phi}) &=&  \frac{\lambda}{4} (\bm{\phi} \, . \, \bm{\phi}  -  \nu^{2})^{2} , \quad \quad \lambda \, , \, \nu > 0  , \label{eq24}
\end{eqnarray}
$ g $ and $ \lambda $ are gauge and scalar coupling constants, respectively, and the constant $ \nu $ is the scalar field vacuum expectation value.

The field equations corresponding to the Georgi-Glashow model are
\begin{eqnarray}
 \triangledown_{\nu} \textbf{F}^{\mu\nu} &=& g \bm{\phi}  \times   \triangledown^{\mu} \bm{\phi}   , \nonumber\\ 
\triangledown_{\mu} \triangledown^{\mu} \bm{\phi}  &=&  - \lambda \bm{\phi}  (\bm{\phi} \, . \, \bm{\phi}  -  \nu^{2}) . \label{eq25}
\end{eqnarray}
The following conditions, low-energy constraints on the classical fields, minimize the energy:
\begin{eqnarray}
\triangledown_{\mu} \bm{\phi} &=&  0  , \label{eq26}  \\   
\bm{\phi} \, . \, \bm{\phi}  &=&  \nu^{2} ,  \label{eq27}  \\  
\textbf{F}_{\mu\nu}  &=&  0 . \label{eq28}
\end{eqnarray}
Since, we want $ \textbf{F}_{\mu\nu} \neq 0 $, we only consider Eq. (\ref{eq26}) and Eq. (\ref{eq27}) and refer to them as vacuum conditions.
Notice that these conditions are the same as the 't Hooft-Polyakov monopole solution constraints in the boundary.
In Sec. \ref{sec4} we generalize these vacuum conditions for the bulk as well as the boundary.

the Higgs field $ \bm{\phi} $ is a vector in color space. Therefore, it has a magnitude and a direction and can be written as
\begin{equation}
\bm{\phi} =   \phi \, \textbf{n}  \, , \quad  (\textbf{n} \, . \, \textbf{n} = 1) , \label{eq29}
\end{equation}
where $ \phi $ is the magnitude (and has the unit and dimension) of $ \bm{\phi} $, and $ \textbf{n} $ is a dimensionless unit vector with a unity
magnitude having the direction of $ \bm{\phi} $.
From Eq. (\ref{eq29}) one gets
\begin{equation}
\triangledown_{\mu} \bm{\phi} =  (\partial_{\mu} \phi) \textbf{n} + \phi \triangledown_{\mu} \textbf{n}     , \label{eq30}
\end{equation}
where
\begin{eqnarray}
 \triangledown_{\mu} \textbf{n} &=&  \partial_{\mu} \textbf{n} + g \textbf{A}_{\mu} \times \textbf{n}  , \nonumber  \\   
&\Rightarrow& \textbf{n} \times \triangledown_{\mu} \textbf{n} = \textbf{n} \times  \partial_{\mu} \textbf{n}
+g \textbf{A}_{\mu} - g (\textbf{A}_{\mu} . \textbf{n}) \textbf{n} , \nonumber  \\  
&\Rightarrow& \textbf{A}_{\mu} =  (\textbf{A}_{\mu} . \textbf{n}) \textbf{n} + \frac{1}{g} \partial_{\mu} \textbf{n} \times \textbf{n} + 
\frac{1}{g} \textbf{n} \times \triangledown_{\mu} \textbf{n}  . \label{eq31}
\end{eqnarray}
Introducing two new fields, $ A_{\mu} $ and $ \textbf{X}_{\mu} $, so that
\begin{eqnarray}
A_{\mu} &=&  \textbf{A}_{\mu} . \textbf{n}  , \nonumber  \\   
\textbf{X}_{\mu} &=&  \frac{1}{g} \textbf{n} \times \triangledown_{\mu} \textbf{n} \, , \quad  (\textbf{X}_{\mu} \, . \, \textbf{n} = 0) , \label{eq32}
\end{eqnarray}
we have
\begin{equation}
\textbf{A}_{\mu} = A_{\mu}  \textbf{n} + \frac{1}{g} \partial_{\mu} \textbf{n} \times \textbf{n} +  \textbf{X}_{\mu}   , \label{eq33}
\end{equation}
which is nothing but the Cho extended decomposition.

There are 15 [3 (for $ \bm{\phi} $) + 12 (for $ \textbf{A}_{\mu} $)]   off-shell degrees of freedom in the SU(2) Georgi-Glashow model.
According to Eq. (\ref{eq29}) and Eq. (\ref{eq33}), we have proposed four new fields, ($ \phi $, $ \textbf{n} $, $ A_{\mu} $, $ \textbf{X}_{\mu} $) .
We rewrite the Higgs field $ \bm{\phi} $ and Yang-Mills field $ \textbf{A}_{\mu} $ based on these new fields:
\begin{eqnarray}
\bm{\phi} &=&   \phi \, \textbf{n}  , \nonumber  \\    
\textbf{A}_{\mu} &=& A_{\mu}  \textbf{n} + \frac{1}{g} \partial_{\mu} \textbf{n} \times \textbf{n} +  \textbf{X}_{\mu} , \label{eq34}
\end{eqnarray}
where there are some constraints:
\begin{equation}
\textbf{n} \, . \, \textbf{n}  = 1 \, , \quad \quad \textbf{n} \, . \, \textbf{X}_{\mu} =0. \label{eq35}
\end{equation}
Notice that there are still 15 [1 (for $ \phi $) + 2 (for $ \textbf{n} $) + 4 (for $ A_{\mu} $) + 8 (for $ \textbf{X}_{\mu} $)] off-shell degrees of freedom.
So the off-shell degrees of freedom are unchanged.

Substituting new variables (\ref{eq34}) in the Georgi-Glashow equations (\ref{eq25}), we get
\begin{eqnarray}
&& \triangledown_{\nu} \textbf{F}^{\mu\nu} = g^{2}  \phi^{2} \, \textbf{X}^{\mu}   , \label{eq36} \\ 
&& (\partial_{\mu} \partial^{\mu} \phi) \textbf{n} + 2 g (\partial_{\mu} \phi) (\textbf{X}^{\mu} \times \textbf{n})
+g \phi \triangledown_{\mu} (\textbf{X}^{\mu} \times \textbf{n}) = - \lambda \phi  (\phi^{2}  -  \nu^{2}) \textbf{n} . \label{eq37}
\end{eqnarray}
Equation (\ref{eq37}) can be decomposed to two equations:
\begin{eqnarray}
\partial_{\mu} \partial^{\mu} \phi &=& g^{2} \phi \, \textbf{X}_{\mu} \, . \, \textbf{X}^{\mu}  - \lambda \phi  (\phi^{2}  -  \nu^{2}) , \label{eq38} \\ 
\triangledown_{\mu} [\phi^{2} \textbf{X}^{\mu}] &=& 0 .  \label{eq39}
\end{eqnarray}
Note that Eq. (\ref{eq39}) can be obtained from Eq. (\ref{eq36}):
\begin{equation}
 \triangledown_{\nu} \textbf{F}^{\mu\nu} = g^{2}  \phi^{2} \, \textbf{X}^{\mu} \quad \Rightarrow \quad 
\triangledown_{\mu} \triangledown_{\nu} \textbf{F}^{\mu\nu} = g^{2} \triangledown_{\mu} \,  [\phi^{2} \textbf{X}^{\mu}] = 0   . \label{eq40}
\end{equation}
So there are just two independent equations:
\begin{eqnarray}
 \triangledown_{\nu} \textbf{F}^{\mu\nu} &=& g^{2}  \phi^{2} \, \textbf{X}^{\mu} , \nonumber  \\ 
\partial_{\mu} \partial^{\mu} \phi &=& g^{2} \phi \, \textbf{X}_{\mu} \, . \, \textbf{X}^{\mu}  - \lambda \phi  (\phi^{2}  -  \nu^{2}) .  \label{eq41}
\end{eqnarray}
These equations can be derived from the following Lagrangian,
\begin{eqnarray}
L &=& \frac{1}{2}  (\partial_{\mu} \phi) (\partial^{\mu} \phi) + \frac{1}{2} g^{2} \phi^{2} \textbf{X}_{\mu} \, . \, \textbf{X}^{\mu} , \nonumber \\
& & - \frac{1}{4} \textbf{F}_{\mu\nu} \, . \, \textbf{F}^{\mu\nu} - \frac{\lambda}{4} (\phi^{2}  -  \nu^{2})^{2}   , \label{eq42}
\end{eqnarray}
which is reformulated as the Georgi-Glashow Lagrangian.

the Euler-Lagrange equations for new variables $ A_{\mu} $, $ \textbf{X}_{\mu} $, and $ \phi $ are
\begin{eqnarray}
\textbf{n} \, . \, \triangledown_{\nu} \textbf{F}^{\mu\nu} &=& 0 , \label{eq43} \\ 
\triangledown_{\nu} \textbf{F}^{\mu\nu} &=& g^{2}  \phi^{2} \, \textbf{X}^{\mu} , \label{eq44}  \\
\partial_{\mu} \partial^{\mu} \phi &=& g^{2} \phi \, \textbf{X}_{\mu} \, . \, \textbf{X}^{\mu}  - \lambda \phi  (\phi^{2}  -  \nu^{2}) ,  \label{eq45}
\end{eqnarray}
and variation with respect to $ \textbf{n} $ does not lead to a new equation and it gets a trivial identity. 
Moreover, considering the constraint (\ref{eq35}), Eq. (\ref{eq43}) can be derived from Eq. (\ref{eq44}).

Notice that the equations of motion of the reformulated Georgi-Glashow model Eqs. (\ref{eq44}) and  (\ref{eq45}) 
  are the same as the original ones Eq.  (\ref{eq41}). Hence, our reformulation does not change the dynamics of the Georgi-Glashow model,
at least at the classical level. In the next section, we study two vacuum conditions for the reformulated Georgi-Glashow model. We also
study the constraint of no external charge, and we see that this constraint leads to the Cho extended Lagrangian of the SU(2) Yang-Mills theory.
In addition, we investigate the condensate phase of the reformulated Georgi-Glashow model, and we reach a
generalization of the Cho-Faddeev-Niemi Lagrangian.

\section{constraints on the classical fields of the Georgi-Glashow model} \label{sec4}
This section is devoted to some constraints on the classical fields of reformulated Georgi-Glashow model.
First, we consider vacuum condition (\ref{eq26}), and we show that it leads to the Cho restricted theory.
Then we study the constraint of no external charge. This constraint leads to either the Cho extended Lagrangian of the SU(2) Yang-Mills theory or a Yang-Mills-Higgs theory
in which the Higgs field and the Yang-Mills field are decoupled and the Yang-Mills part is the same as in the Cho restricted theory.
Finally, the other vacuum condition (\ref{eq27}) will be considered. In this condensate phase, we derive a Lagrangian that is a generalization
of the Cho-Faddeev-Niemi Lagrangian.

\subsection{The vacuum condition which leads to the Cho restricted decomposition}  \label{subsec4.1}

Consider the vacuum condition (\ref{eq26}) :
\begin{eqnarray}
&& \triangledown_{\mu} \bm{\phi} = (\partial_{\mu} \phi)  \textbf{n} + \phi \triangledown_{\mu} \textbf{n} = 0 , \nonumber \\ 
&& \Rightarrow \quad \textbf{n} \, . \, [(\partial_{\mu} \phi)  \textbf{n} + \phi \triangledown_{\mu} \textbf{n} ] = 0, \nonumber \\
&& \Rightarrow \partial_{\mu} \phi =0 \quad \Rightarrow \quad \phi = constant.  \label{eq46}
\end{eqnarray}
On the other hand, from Eq. (\ref{eq32}), we get
\begin{equation}
\textbf{X}_{\mu} =  \frac{1}{g} \textbf{n} \times \triangledown_{\mu} \textbf{n} \quad \Rightarrow \quad 
\triangledown_{\mu} \textbf{n} = g \textbf{X}_{\mu}  \times \textbf{n}. \label{eq47}
\end{equation}
Eqs. (\ref{eq46}) and (\ref{eq47}) implicate $ \textbf{X}_{\mu} = 0 $. Therefore, for this vacuum condition we have
\begin{equation}
\phi = constant \, , \quad \textbf{A}_{\mu} =  \widehat{\textbf{A}}_{\mu} = A_{\mu}  \textbf{n} + \frac{1}{g} \partial_{\mu} \textbf{n} \times \textbf{n} , \label{eq48}
\end{equation}
which is the Cho restricted decomposition. Our reformulated Georgi-Glashow Lagrangian in this case is
\begin{equation}
L =  - \frac{1}{4} \widehat{\textbf{F}}_{\mu\nu} \, . \, \widehat{\textbf{F}}^{\mu\nu}  + constant  , \label{eq49}
\end{equation}
and the equation of motion is
\begin{equation}
 \widehat{\triangledown}_{\mu} \widehat{\textbf{F}}^{\mu\nu} = 0 . \label{eq50}
\end{equation}

\subsection{No external source condition which leads to Cho extended decomposition}  \label{subsec4.2}

Suppose that there is no external charge in the reformulated Georgi-Glashow model:
\begin{equation}
\triangledown_{\nu} \textbf{F}^{\mu\nu} = g^{2}  \phi^{2} \, \textbf{X}^{\mu} = 0 . \label{eq51}
\end{equation}
There are two options for satisfying the above equation:
one is $  \phi =0 $ and the other $  \textbf{X}^{\mu} = 0 $. The first one, $  \phi =0 $, leads to the Cho extended SU(2) Yang-Mills Lagrangian.
In this case, we have
\begin{eqnarray}
\bm{\phi} &=&   \phi \, \textbf{n} = 0 , \nonumber  \\    
\textbf{A}_{\mu} &=& A_{\mu}  \textbf{n} + \frac{1}{g} \partial_{\mu} \textbf{n} \times \textbf{n} +  \textbf{X}_{\mu}  , \label{eq52}
\end{eqnarray}
and the Lagrangian is
\begin{eqnarray}
L  &=& - \frac{1}{4} \widehat{\textbf{F}}_{\mu\nu} \, . \, \widehat{\textbf{F}}^{\mu\nu} - \frac{1}{4} ( \widehat{\triangledown}_{\mu} \textbf{X}_{\nu} -  \widehat{\triangledown}_{\nu} \textbf{X}_{\mu}).( \widehat{\triangledown}^{\mu} \textbf{X}^{\nu} -  \widehat{\triangledown}^{\nu} \textbf{X}^{\mu}) \nonumber \\
 & & - \frac{g^{2}}{4} (\textbf{X}_{\mu} \times \textbf{X}_{\nu}).(\textbf{X}^{\mu} \times \textbf{X}^{\nu}) 
 - \frac{g}{2} \widehat{\textbf{F}}_{\mu\nu} \, . \, (\textbf{X}^{\mu} \times \textbf{X}^{\nu})   + constant . \label{eq53} 
\end{eqnarray}
The second case, $  \textbf{X}^{\mu} = 0 $, leads to a Yang-Mills-Higgs theory
in which the Higgs field and the Yang-Mills field are decoupled, and the Yang-Mills part is the same as in the Cho restricted theory:
\begin{equation}
L = \frac{1}{2}  (\partial_{\mu} \phi) (\partial^{\mu} \phi)  - \frac{\lambda}{4} (\phi^{2}  -  \nu^{2})^{2}
 - \frac{1}{4} \widehat{\textbf{F}}_{\mu\nu} \, . \, \widehat{\textbf{F}}^{\mu\nu}   . \label{eq54}
\end{equation}
the Euler-Lagrange equations for this case are
\begin{eqnarray} 
\triangledown_{\nu} \widehat{\textbf{F}}^{\mu\nu} &=& 0 , \nonumber  \\
\partial_{\mu} \partial^{\mu} \phi &=&   - \lambda \phi  (\phi^{2}  -  \nu^{2}) ,  \label{eq55}
\end{eqnarray}
which are two second-order decoupled differential equations. Therefore, the condition $ \textbf{X}_{\mu} \neq 0 $ is essential for interaction
between the Higgs field and the Yang-Mills field.

\subsection{Condensate phase}  \label{subsec4.3}

Vacuum condition (\ref{eq27}) in which the Higgs field takes the vacuum expectation value, $ \phi = \nu $, leads to the following effective Lagrangian for the condensate phase:
\begin{eqnarray}
L &=&  \frac{1}{2} g^{2} \nu^{2} \, \textbf{X}_{\mu} \, . \, \textbf{X}^{\mu} - \frac{1}{4} \textbf{F}_{\mu\nu} \, . \, \textbf{F}^{\mu\nu} \nonumber \\
 &=&  \frac{1}{2} g^{2} \nu^{2} \, \textbf{X}_{\mu} \, . \, \textbf{X}^{\mu} - \frac{1}{4} \widehat{\textbf{F}}_{\mu\nu} \, . \, \widehat{\textbf{F}}^{\mu\nu} \nonumber \\
 & &  - \frac{1}{4} ( \widehat{\triangledown}_{\mu} \textbf{X}_{\nu} -  \widehat{\triangledown}_{\nu} \textbf{X}_{\mu}).( \widehat{\triangledown}^{\mu} \textbf{X}^{\nu} -  \widehat{\triangledown}^{\nu} \textbf{X}^{\mu}) \nonumber \\
 & & - \frac{g^{2}}{4} (\textbf{X}_{\mu} \times \textbf{X}_{\nu}).(\textbf{X}^{\mu} \times \textbf{X}^{\nu}) 
 - \frac{g}{2} \widehat{\textbf{F}}_{\mu\nu} \, . \, (\textbf{X}^{\mu} \times \textbf{X}^{\nu})     . \label{eq56}
\end{eqnarray}

In this phase, $ \textbf{X}_{\mu} $ is massive and its mass is
\begin{equation}
m_{\textbf{X}} = g \, \nu . \label{eq57}
\end{equation}
Considering Eq.  (\ref{eq39}), in this case we have
\begin{equation}
\triangledown_{\mu}  \textbf{X}^{\mu} = 0 \quad \Rightarrow \quad \widehat{\triangledown}_{\mu}  \textbf{X}^{\mu} = 0 .  \label{eq58}
\end{equation}
This condition was imposed on the Cho extended decomposition in order to compensate for the two extra degrees introduced by $ \textbf{n} $.
In the Cho decomposition, there are
14 [2 (for $ \textbf{n} $) + 4 (for $ A_{\mu} $) + 8 (for $ \textbf{X}_{\mu} $)] off-shell degrees of freedom, while there are 12 off-shell degrees of freedom for an SU(2)
Yang-Mills field. The field $ \textbf{n} $ is responsible for these two extra degrees, and it has led people to search for
two extra constraints that can demolish these two extra degrees created by $ \textbf{n} $ \cite{FN1,FN2,FN3,FN4,Shabanov1,Shabanov2,Gies}.
These ideas have been criticized and discussed in \cite{Cho4}.
In our reformulation this problem does not occur and as we mentioned before, both the Georgi-Glashow model and our reformulation of this model have 15 degrees.

Finally, we generalize the Faddeev-Niemi Lagrangian for the condensate phase. Substituting Eqs. (\ref{eq15}) and (\ref{eq19}) in Lagrangian (\ref{eq56}),
we get
\begin{eqnarray}
L &=& \frac{1}{2} \frac{\nu^{2}}{g^{2}} \, \varphi ^{\ast} \varphi \, \,   \partial_{\mu} \textbf{n} \, . \,  \partial^{\mu} \textbf{n}  \nonumber\\ 
& & -\frac{1}{4} F_{\mu\nu} F^{\mu\nu} + \frac{1}{2g^{4}} ( \partial_{\mu} \textbf{n}.\partial_{\nu} \textbf{n} - \eta_{\mu\nu} \partial_{\lambda} \textbf{n}.\partial^{\lambda} \textbf{n} )
(D^{\mu}\varphi)^{\ast}  (D^{\nu}\varphi) \nonumber\\
&& + \frac{i}{2g^{3}} H_{\mu\nu} (D^{\mu}\varphi)^{\ast}  (D^{\nu}\varphi)  - \frac{1}{2} H_{\mu\nu} F^{\mu\nu} (1-\frac{\varphi ^{\ast} \varphi}{g^{2}})  \nonumber\\
&& - \frac{1}{4}  H_{\mu\nu} H^{\mu\nu} (1-\frac{\varphi ^{\ast} \varphi}{g^{2}})^{2} ,  \label{eq59}
\end{eqnarray}
where the first term is added to the Faddeev-Niemi Lagrangian and it leads to new results.
In an upcoming paper, we show how the Skyrme-Faddeev Lagrangian can be derived from the above Lagrangian by considering some more constraints on the classical Faddeev-Niemi variables.
Therefore, the Skyrme-Faddeev Lagrangian which describes knotlike solitons can be interpreted as an effective Lagrangian of the condensate phase of
our reformulation of the Georgi-Glashow model.

\section{Conclusion} \label{sec5}

the Lagrangian based on the new variables is a method that is used by many authors for different purposes.
The strong motivation for this method is that by using these variables one can reach a theory more appropriate for the low-energy limits of the original Lagrangian.
We take a new look at Cho-Faddeev-Niemi decomposition and their proposed variables for the SU(2) Yang-Mills field.
Their new variables and their decomposition will be more comprehensible if we consider a reformulation of the Georgi-Glashow model that is proposed in this paper.
For example, according to our reformulation both the Cho restricted and the extended SU(2) Yang-Mills Lagrangian are special cases of the reformulated Georgi-Glashow model with some constraints on the classical fields that we refer to them as vacuum conditions.
Furthermore, we get two new limits in this model.
In one of them, the constraint of no external charge leads to a Yang-Mills-Higgs theory in which the Higgs field is decoupled from the Yang-Mills field.
In this limit, the Yang-Mills field decomposition is the same as the Cho restricted decomposition.
In the other limit, the condensate phase, the valence part of Yang-Mills field, $ \textbf{X}_{\mu} $, becomes massive.
This limit leads to a generalization of the Cho-Faddeev-Niemi Lagrangian in which a new mass term for the field $ \textbf{X}_{\mu} $ is included.
In upcoming work, we show that one can reach the Skyrme-Faddeev Lagrangian by considering one more decomposition.
So we will show that the Skyrme-Faddeev theory of the nonlinear sigma model and our reformulation of the Georgi-Glashow model with some extra constraints on the fields have identical topological structures.

\section{\boldmath Acknowledgments}
I am grateful to S. Deldar for stimulating and fruitful discussions.

\end{document}